\documentclass[onecolumn,noshowpacs,nopreprintnumbers,noamsmath,noamssymb,eqsecnum,noshowkeys,floatfix,prc]{revtex4}

\usepackage{graphicx}
\usepackage{dcolumn}
\usepackage{bm}

\begin{document}

\newcommand{\vcr}{\mbox{${\bf r}\,$}}
\newcommand{\vcj}{\mbox{${\bf j}\,$}}
\newcommand{\vcJ}{\mbox{${\bf J}\,$}}
\newcommand{\cH}{\mbox{$\cal{H}$}}
\newcommand{\cN}{\mbox{$\cal{N}$}}
\newcommand{\nh}{\mbox{$\hat{n}$}}
\newcommand{\Ph}{\mbox{$\hat{P}$}}
\newcommand{\kh}{\mbox{$\hat{k}$}}
\newcommand{\sh}{\mbox{$\hat{\sigma}$}}
\newcommand{\vchk}{\mbox{$\hat{\bf k}$}}
\newcommand{\nbh}{\mbox{$\hat{\bar{n}}$}}
\newcommand{\pvh}{\mbox{$\hat{\vec{p}}$}}
\newcommand{\magsqr}[1]{\mbox{$\left|{#1}\right|^2$}}
\newcommand{\magabs}[1]{\mbox{$\left|{#1}\right|$}}
\newcommand{\nedots}{\mbox{$.\cdot^{\textstyle{.}}$}}


\title{Nuclear structure near the neutron dripline: lattice HFB \\  
      calculations with high-energy continuum coupling\\}

\author{V.E. Oberacker}%
 \email{volker.e.oberacker@vanderbilt.edu}
\author{A.S. Umar}
 \email{umar@compsci.cas.vanderbilt.edu}
\author{E. Ter\'an}
 \email{edgar.teran@vanderbilt.edu}
\affiliation{%
Department of Physics and Astronomy, Vanderbilt University, 
Nashville, TN 37235, USA\\
}%

\date{\today}
\begin{abstract}
We have developed a new Hartree-Fock-Bogoliubov (HFB) code which has been
specifically designed to study
ground state properties of nuclei near the neutron and
proton drip lines. The unique feature of our code is that it takes
into account the strong coupling to high-energy continuum states,
up to an equivalent single-particle energy of 60 MeV. We solve the HFB
equations for deformed, axially symmetric even-even nuclei in 
coordinate space on a 2-D lattice with Basis-Spline methods. 
For the p-h channel, the Skyrme (SLy4) effective N-N interaction is utilized, 
and for the p-p and h-h channel we use a delta interaction. We present results for
binding energies, deformations, normal densities and pairing densities, Fermi levels,
and pairing gaps. In particular, we will discuss neutron-rich isotopes of oxygen
($^{22}O$) and tin ($^{150}Sn$).
\end{abstract}
\pacs{21.60 Jz, 24.30 Cz}
\keywords{Nuclear structure, HFB, Basis-splines}
\maketitle
\section{\label{sec:intro}Introduction\protect\\}
One of the fundamental questions of nuclear structure physics is: what are the limits
of nuclear stability? How many neutrons or protons can we add to a given nucleus
before it becomes unstable against spontaneous neutron or proton emission? If one connects
the isotopes with zero neutron separation energy, $S_n=0$, in the nuclear chart one obtains the
neutron dripline. Similarly, the proton dripline is defined by the condition $S_p=0$. 
Another limit to stability is the superheavy element region around $Z=124-126$ and $N=184$
which is formed by a delicate balance
between strong Coulomb repulsion and additional binding due to closed shells. 
The nuclear chart shows less than 300 stable nuclear isotopes, and about 1700 additional
isotopes have been synthesized and studied in accelerator experiments.
Nuclei in between the proton and neutron driplines are unstable against $\beta$-decay.
Nuclei outside the driplines decay by spontaneous neutron emission or 
proton radioactivity. The neutron-rich side, in particular, exhibits thousands of
nuclear isotopes still to be explored (`terra incognita').

Some of these exotic nuclei can be studied with existing first-generation Radioactive
Ion Beam Facilities (e.g. HRIBF at Oak Ridge, NSCL at Michigan State
University, GANIL in France, GSI in Germany, and RIKEN in Japan). Several
countries are planning to construct new `second generation' RIB facilities,
in particular for the exploration of neutron rich isotopes.
In the United States, the DOE/NSF Long Range Plan for Nuclear Physics, published in April
2002, gives the highest priority for new construction to RIA (Rare
Isotope Accelerator). RIA is a bold new concept in exotic beam
facilities in that it combines both of the known rare isotope production
techniques: the ISOL method (thick-target spallation) and high-energy projectile
fragmentation.

\begin{figure}[ht]
\begin{center}
\includegraphics[scale=0.5]{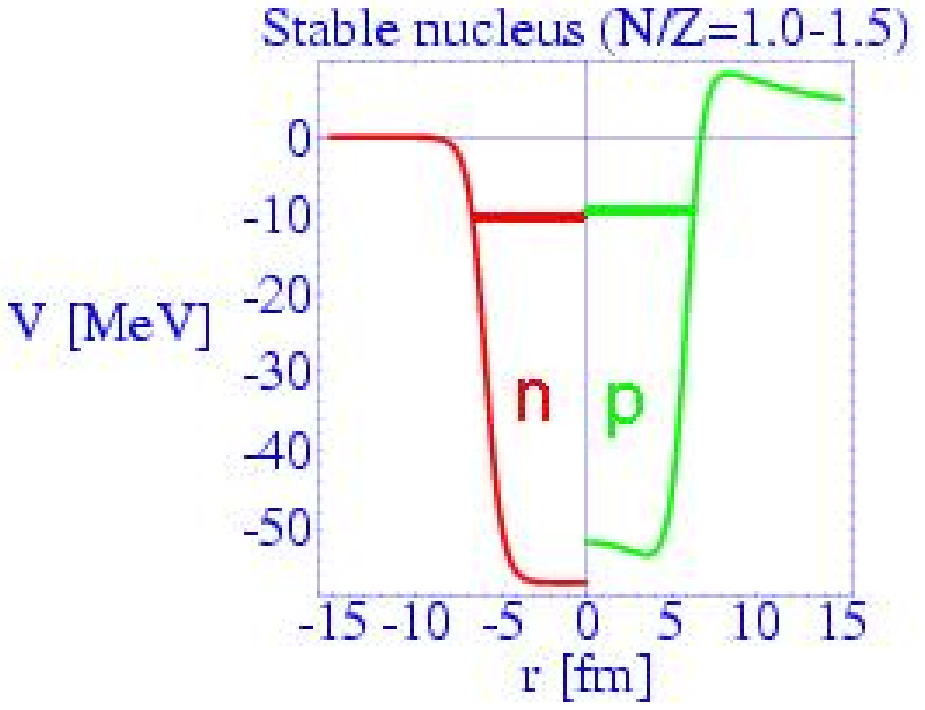}
\includegraphics[scale=0.54]{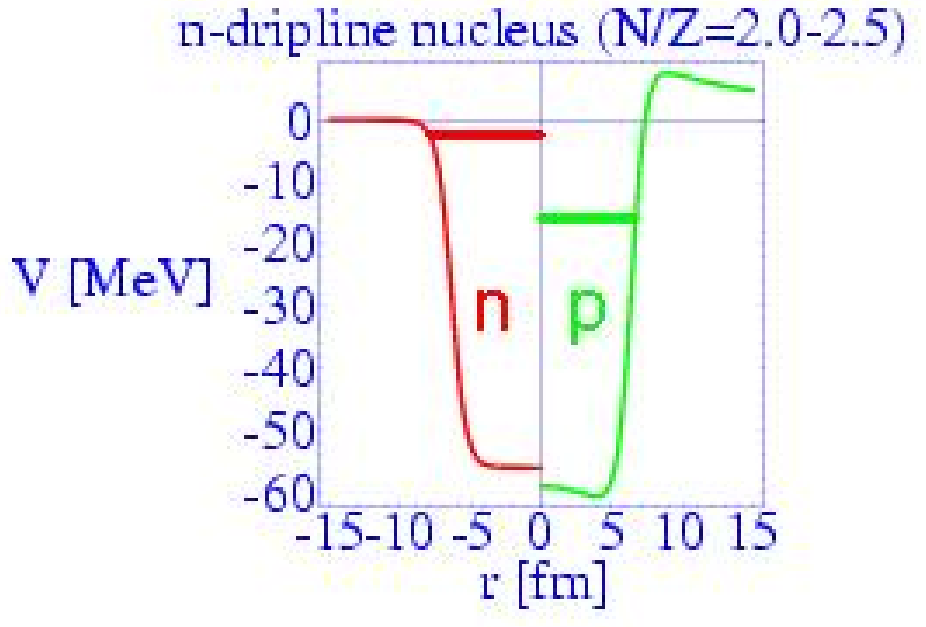}
\end{center}
\caption{One-body `mean field' potential for protons and neutrons. Left: for
a nucleus near the stability line. Right: for a nucleus near the neutron
dripline. The Fermi levels for protons and neutrons are indicated by
horizontal lines.}
\label{sketch_pot}
\end{figure}

Theories predict profound differences between the known isotopes
near stability and the exotic nuclei at the driplines \cite{ISOL97,RIA1}:
for n-rich nuclei, as the Fermi level approaches the particle continuum at $E=0$
(see Fig.~\ref{sketch_pot}) weakly bound states couple strongly to
the continuum states giving rise to neutron halos and neutron skins. 
Theories also expect large pairing correlations and new collective modes (e.g. `pigmy
resonance'), a weakening of the spin-orbit force leading to a quenching of the shell gaps,
and perhaps new magic numbers.

Furthermore, RIA will allow us to address fundamental questions in nuclear
astrophysics: more than half of all elements heavier
than iron are thought to be produced in supernovae explosions
by the rapid neutron capture process (r-process). The r-process
path contains many exotic neutron-rich nuclei which can only
be studied with RIA. Also, the predicted neutron skins would allow
us to measure the properties of pure neutron matter which is of great
interest for the study of neutron stars.

These experimental developments as well as recent advances in
computational physics have sparked renewed interest in nuclear
structure theory. There are several types of approaches in nuclear structure theory
\cite{RIA1}: for the lightest nuclei, ab-initio calculations
(Green's function Monte Carlo, no-core shell model) based
on the bare N-N interaction are possible \cite{NB97}. Medium-mass nuclei up to
$A \sim 60$ may be treated in the large-scale shell model approach
\cite{KDL97}. For heavier nuclei one utilizes either nonrelativistic
\cite{DF84,DN96} or relativistic \cite{PV97,ND96} mean
field theories. 


\section{HFB theory in coordinate space: deformed nuclei on a 2-D lattice\protect\\}

An accurate treatment of the pairing interaction is essential for
exotic nuclei \cite{DF84,DN96}.
As we move away from the valley of stability, surprisingly little is
known about the pairing force: for example, what is its density
dependence? Neutron-rich nuclei are expected to be highly superfluid
due to continuum excitation of neutron `Cooper pairs'. These
large pairing correlations near the driplines can no longer be described
by a small residual interaction. It becomes necessary to treat the
mean field and the pairing field in a single self-consistent
theory, known as the Hartree-Fock-Bogoliubov theory (HFB). Near the
neutron dripline, the Fermi energy approaches $E=0$
and the outermost nucleons are weakly bound (which implies a
large spatial extent); they are strongly coupled to the particle
continuum at $E>0$. These features represent major challenges for the
numerical solution.

Most HFB calculations to date are carried out in a truncated discrete harmonic
ocillator basis, see e.g. \cite{RS80,PS90,CB92,ER93,SD00}.

This approach is quite appropriate for nuclei in the vicinity of the stability line.
However, farther from stability, the continuum states become important and 
coordinate-based representations have numerous advantages: for example, the
well-known `French code' uses a truncated 3-D Hartree-Fock
basis \cite{TH96} which consists of both localized states and
discretized continuum states; however, in this approach one can only 
include continuum states up to about $5$ MeV of excitation energy.
For nuclei in the vicinity of the driplines, continuum states with an equivalent
single-particle energy of up to 60 MeV must be taken into account. 
One-dimensional calculations for spherical nuclei have been carried out
in coordinate space for many years \cite{DF84,DN96}, but only recently has
our Vanderbilt group succeeded in generalizing this approach to the important
case of deformed axially symmetric nuclei (HFB on a 2-D lattice)
\cite{OU01,TOU02a,TOU02b}.
We utilize a novel computational technique, a Basis-Spline representation of
wavefunctions and operators, which allows us to accurately describe
high-energy continuum states in two space dimensions $(z,r)$.

The many-body Hamiltonian in occupation number representation has the form
\begin{equation}
\hat{H} = \sum_{i,j} < i|\ t\ |j> \ \hat{c}_i^\dagger \ \hat{c}_j
+ \frac{1}{2} \sum_{i,j,k,l} <ij|\ v^{(2)} \ |kl>
\ \hat{c}_i^\dagger \ \hat{c}_j^\dagger \ \hat{c}_l \ \hat{c}_k 
+ \frac{1}{6} \sum_{i,j,k,l,m,n} <ijk|\ v^{(3)} \ |lmn>
\ \hat{c}_i^\dagger \ \hat{c}_j^\dagger \ \hat{c}_k^\dagger
\ \hat{c}_n \ \hat{c}_m  \ \hat{c}_l \ .
\end{equation}
The general linear transformation from particle operators $\hat{c},\hat{c}^\dagger$ 
to quasiparticle operators $\hat{\beta},\hat{\beta}^\dagger$ take the form \cite{RS80}:
\begin{equation}
\left( 
\begin{array}{c}
\hat{\beta} \\ 
\hat{\beta}^\dagger 
\end{array}
\right) =
\left( 
\begin{array}{cc}
 U^\dagger & V^\dagger \\ 
  V^T & U^T 
\end{array}
\right) 
\left( 
\begin{array}{c}
\hat{c} \\ 
\hat{c}^\dagger
\end{array}
\right) \ .
\end{equation}
\noindent The HFB approximate ground state of the many-body system is defined as a
vacuum with respect to quasiparticles
\begin{equation}
\hat{\beta}_k \ | \Phi_0 > \ = \ 0 \ .
\end{equation}
\noindent The HFB ground state energy including the constraint on the 
particle number $N$ is given by
\begin{equation}
E({\mathcal{R}}) = < \Phi_0 | \hat{H} - \lambda \hat{N} | \Phi_0 >
\end{equation}
\noindent where the Lagrange multiplier $\lambda$ is the Fermi energy of the system.
The equations of motion are derived from the variational principle
\begin{equation}
\delta |_{\mathcal{R}} \ [ E( {\mathcal{R}} ) - {\rm{tr}}\ 
 \Lambda ( {\mathcal{R}}^2 - {\mathcal{R}} ) ]  =  0 \ ,
\end{equation}
\noindent where $\mathcal{R}$ represents the generalized density matrix.


\subsection{Quasiparticle wave functions and densities}

In practice, it is to convenient to transform the standard HFB equations into a coordinate 
space representation and solve the resulting differential equations on a lattice. For this 
purpose, one defines two types of quasiparticle wavefunctions $\phi_1$ and $\phi_2 $ 
\begin{equation}
\phi_1^* (E_\alpha, {\bf r} \sigma q) \ = \ \sum_i U_{i\alpha} \ (2 \sigma) \ \phi_i ({\bf r} -\sigma q) ,
\ \ \ \ \ \ 
\phi_2 (E_\alpha, {\bf r} \sigma q) \ = \ \sum_i V_{i\alpha}^* \ \phi_i ({\bf r} \sigma q) \ .
\end{equation}
The basis wavefunctions $\phi_i$ depend on the
position vector ${\bf r}$, the spin projection $\sigma = \pm \frac{1}{2}$, and 
the isospin projection $q$ ($q=+\frac{1}{2}$ corresponds to protons and $q=-\frac{1}{2}$ to
neutrons). \\

From these wavefunctions we obtain the following expressions for the normal density 
$\rho_q({\bf r})$ and the pairing density $\tilde \rho_q({\bf r})$
\begin{equation}
\rho_q({\bf r}) = \sum_{\sigma} \sum_{\alpha} \phi_{2,\alpha} ({\bf r} \sigma q)
                  \ \phi_{2,\alpha}^* ({\bf r} \sigma q) , \ \ \ \ \ \ 
\tilde{\rho_q}({\bf r}) = - \sum_{\sigma} \sum_{\alpha} \phi_{2,\alpha} ({\bf r} \sigma q)
                           \ \phi_{1,\alpha}^* ({\bf r} \sigma q) \ . 
\end{equation}
The quasiparticle energy $E_\alpha$ is denoted by index $\alpha$ for simplicity.
In principle, the sums go over all the energy states, but in practice a cutoff 
is introduced (see later). The physical interpretation of $\tilde{\rho_q}$
has been discussed in \cite{DN96}:
the quantity $[\tilde{\rho_q}({\bf r})\ \Delta V /2]^2$ gives the probability to find a
\emph{correlated} pair of nucleons with opposite spin projection in the volume
element $\Delta V$. The kinetic energy density $\tau_q({\bf r})$ is found to be
\begin{equation}
\tau_q({\bf r}) = \sum_{\sigma} \sum_{\alpha} | \nabla \ \phi_{2,\alpha} ({\bf r} \sigma q) |^2 \ .
\label{eq:kinetic_energy}
\end{equation}

\subsection{Binding energy functional}

In our calculations we utilize the Skyrme two-body effective N-N interaction
\begin{eqnarray}
v^{(2)}_{12} &=& t_{0} \ (1+x_{0}\Ph_{\sigma }) \ \delta ({\bf r}_{1}-{\bf r}_{2})
+ \frac{1}{2} \ t_{1} \ (1+x_{1}\Ph_{\sigma }) \ \{ \delta ({\bf r}_{1}-{\bf r}_{2})\kh^{2} ) + 
\kh'^{2}\delta ({\bf r}_{1}-{\bf r}_{2})\} \nonumber\\ 
&+& t_{2} \ (1+x_{2}\Ph_{\sigma })\ \vchk' \cdot \delta ({\bf r}_{1}-{\bf r}_{2})\ \vchk
+ \frac{1}{6} \ t_{3} \ (1+x_{3}\Ph_{\sigma })\ \rho ^{\gamma }\ \delta ({\bf r}_{1}-{\bf r}_{2})
+ i\ W_{0} \ (\sh_{1}+\sh_{2})\cdot \{\vchk'\times\delta ({\bf r}_{1}-{\bf r}_{2})\vchk \} \ .
\end{eqnarray}
The density-dependent term proportional to $t_3$ simulates the three-body N-N interaction
$v^{(3)}$.

The total binding energy of the nucleus 
\begin{equation}
\label{bind}
E_0^{HFB} = \langle \Phi_0 | H | \Phi_0 \rangle =  E_{kin} + E_{Sky} + E_{Sky,LS}
             + E_{Coul} + E_{pair} + E_{cm} \ .
\end{equation}
consists of a kinetic energy term, various contributions from the Skyrme effective
N-N interaction (including a spin-orbit term), Coulomb and pairing energy, and a
center-of mass correction due to the mean-field approximation
\begin{eqnarray}
E_{kin} &=& \int d^3r \frac{\hbar^2}{2m}\ \tau({\bf r}) \nonumber \\
E_{Sky} &=& \int d^3r \left [
            \frac{b_0}{2} \ \rho({\bf r})^2
           + b_1 \rho({\bf r}) \tau({\bf r})
	   - \frac{b_2}{2} \rho({\bf r}) \nabla^2 \rho({\bf r})
           + \frac{b_3}{3} \rho({\bf r})^{\alpha + 2}
	   \right]  \nonumber \\	   
           &+& \int d^3r \sum_q \left [ 
           - \frac{b'_0}{2} \rho_q^2 - b'_1 \rho_q \tau_q
           + \frac{b'_2}{2} \rho_q \nabla^2 \rho_q
           - \frac{b'_3}{3} \rho^{\alpha} \rho_q^2
	    \right] \nonumber \\
E_{Sky,LS} &=& \int d^3r \left [
	      - b_4 \rho {\bf \nabla \cdot J} - b'_4 \sum_q \rho_q ({\bf \nabla \cdot J}_q) \ .
              \right]
\label{eq:bind1}
\end{eqnarray}
The Coulomb energy contains the direct term as well as an exchange
term (in Slater approximation)
\begin{equation}
E_{Coul} = 
\frac{e^2}{2} \int d^3r \int d^3 r' \rho_p(\vcr) \frac{1}{|\vcr - \vcr'|} \rho_p(\vcr') 
 - \frac{3}{4} e^2 \left( \frac{3}{\pi} \right)^{1/3}
\int d^3r \left[ \rho_p(\vcr) \right]^{4/3} \ .
\end{equation}


\subsection{Pairing interaction}

In practice, one tends to use \emph{different} effective N-N interactions for
the p-h / h-p channels and for the p-p / h-h channels. Most pairing calculations
utilize a local pairing interaction of the form
\begin{equation}
V_p({\bf r} \sigma, {\bf r}' - \sigma') \ = \ V_0 \ \delta({\bf r} - {\bf r}')
            \ \delta_{\sigma,\sigma'} \ F({\bf r}) \ .
\end{equation}
This parameterization describes two primary pairing forces: 
a pure delta interaction ($F=1$) that gives rise to {\it volume pairing}, and a 
density dependent delta interaction (DDDI) that gives rise to {\it surface pairing}. 
In the latter case, one uses the following phenomenological ansatz \cite{RD99}
for the factor $F$
\begin{equation}
F({\bf r}) \ = \ 1 - \left( \frac{\rho({\bf r})}{\rho_0}  \right)^\gamma
\end{equation}
where $\rho({\bf r})$ is the mass density. 

The pairing contribution to the nuclear binding energy is then
\begin{equation}
E_{pair} = \frac{V_0}{4} \int d^3 r \sum_q 
   \tilde{\rho}_q^{\ 2} ({\bf r}) F({\bf r}) \ .
\end{equation}
An important related quantity is the average pairing gap for protons and
neutrons which can be calculated from the general expression given in \cite{DF84,DN96}
\begin{equation}
<\Delta_q> \ = \ - \frac{1}{2} \ \frac{V_0}{N_q} \int d^3 r \ 
\tilde{\rho_q}({\bf r}) \ \rho_q({\bf r}) \ F({\bf r}) \ .
\end{equation}
where $N_q$ denotes the number of protons or neutrons. Note that the pairing gap
is a positive quantity because $V_0<0$.


\subsection{\label{sec:hfb_eqns}HFB equations and mean fields in coordinate space}

\begin{figure}[h]
\vspace*{0.0cm}
\includegraphics[scale=0.30]{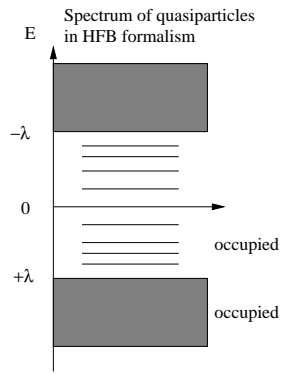}
\vspace*{0.0cm}
\caption{\label{fig:spectrum} Computational challenge in solving the HFB
    equations in coordinate space: the quasiparticle energy spectrum is
    unbounded from above and below.}
\end{figure}

For certain types of effective interactions (e.g. Skyrme mean field and pairing
delta-interactions) the particle Hamiltonian $h$ and the
pairing Hamiltonian $\tilde h$ are diagonal in isospin space and
local in position space, resulting in the following HFB equations with 
a 4x4 structure in spin space:
\begin{equation}
\left( 
\begin{array}{cc}
( h^q -\lambda ) & \tilde h^q \\
\tilde h^q & - ( h^q -\lambda ) 
\end{array}
\right)
\left(
\begin{array}{c}
\phi^q_{1,\alpha} \\  
\phi^q_{2,\alpha}
\end{array}
\right)
 = E_\alpha
\left( 
\begin{array}{c}
\phi^q_{1,\alpha} \\  
\phi^q_{2,\alpha}
\end{array}
\right)
\label{eq:hfbeq2}
\end{equation}
with
\begin{equation}
h^q =
\left( 
\begin{array}{cc}
h^q_{\uparrow \uparrow}({\bf r}) & h^q_{\uparrow \downarrow}({\bf r}) \\
h^q_{\downarrow \uparrow}({\bf r}) & h^q_{\downarrow \downarrow}({\bf r})
\end{array}
\right)
, \ \ \ \ \ \ 
\tilde h^q =
\left( 
\begin{array}{cc}
\tilde h^q_{\uparrow \uparrow}({\bf r}) & \tilde h^q_{\uparrow \downarrow}({\bf r}) \\ 
\tilde h^q_{\downarrow \uparrow}({\bf r}) & \tilde h^q_{\downarrow \downarrow}({\bf r})
\end{array}
\right) \ \ .
\label{hspin}
\end{equation}
The HFB equations have a mathematical structure that is similar to the 
Dirac equation: the spectrum of quasiparticle energies $E$ is unbounded from above {\em
and} below. The spectrum is discrete for $|E|<-\lambda$
and continuous for $|E|>-\lambda$. This is illustrated in Fig. \ref{fig:spectrum}.

As explained in \cite{RS80}, it is forbidden to choose positive and negative
quasiparticle energies at the same time, otherwise it is impossible to satisfy
the anticommutation relations for $\hat{\beta}_\alpha,\hat{\beta}_\alpha^\dagger$.
For even-even nuclei it is customary to solve the HFB equations with a positive
quasiparticle energy spectrum $+E_\alpha$ and consider all negative energy states
as occupied in the HFB ground state.

The HFB mean field Hamiltonian has the same structure as the binding energy functional
\begin{equation}
\label{eq:hf_hamiltonian}
h_q = - {\bf\nabla  \cdot} \frac{\hbar^2}{2 m_q^*({\bf r})} {\bf \nabla} 
    + U_q({\bf r}) + U_{q,LS}({\bf r}) + U_{Coul}({\bf r}) \cdot \delta_{q,p} \ .
\end{equation}
The coordinate - dependent effective mass arises from the densities   
\begin{equation}
\label{effective_mass_def}
\frac{\hbar^2}{2 m_q^*({\bf r})} = \frac{\hbar^2}{2m_q} 
 \ + \ b_1 \ \rho({\bf r})  \ - \ b'_1 \ \rho_q({\bf r}) \ .
\end{equation}
Detailed expressions for the Skyrme mean fields and the Coulomb term are given in
reference \cite{TOU02b}. The DDDI interaction generates the following pairing
mean field for the two isospin orientations $q = \pm \frac{1}{2}$
\begin{equation}
\tilde{h}_q({\bf r}) \ = \ \frac{1}{2} \ V_0 
         \tilde{\rho_q}({\bf r}) F({\bf r}) \ .
\end{equation}


\section{\label{sec:axial_eqns}2-D Reduction for Axially Symmetric Systems}

For simplicity, we assume that the HFB quasiparticle Hamiltonian
is invariant under rotations ${\hat R}_z$ around
the z-axis, i.e. $[{\mathcal{H}},{\hat R}_z]=0$. Due to the axial symmetry
of the problem, it is advantageous to introduce cylindrical coordinates
$(\phi,r,z)$.
It is possible to construct simultaneous
eigenfunctions of the generalized Hamiltonian ${\mathcal{H}}$ and
the z-component of the angular momentum, ${\hat j}_z$
with quantum numbers 
$\Omega = \pm \frac{1}{2}, \pm \frac{3}{2},\pm \frac{5}{2}, ...$ corresponding to
each $nth$ energy state.
The simultaneous quasiparticle eigenfunctions take the form
\begin{eqnarray}
\psi_{n,\Omega,q} (\phi,r,z) = 
\left( 
\begin{array}{c}
\phi^{(1)}_{n,\Omega,q} (\phi,r,z) \\ 
\phi^{(2)}_{n,\Omega,q} (\phi,r,z) \\ 
\end{array} 
\right) =
\frac{1}{\sqrt{2 \pi}}
\left( 
\begin{array}{c}
e^{i(\Omega - \frac 12)\phi} \ \phi^{(1)}_{n,\Omega,q} (r,z,\uparrow) \\ 
e^{i(\Omega + \frac 12)\phi} \ \phi^{(1)}_{n,\Omega,q} (r,z,\downarrow) \\
e^{i(\Omega - \frac 12)\phi} \ \phi^{(2)}_{n,\Omega,q} (r,z,\uparrow) \\ 
e^{i(\Omega + \frac 12)\phi} \ \phi^{(2)}_{n,\Omega,q} (r,z,\downarrow)
\end{array} 
\right) \ .
\label{eq:wvfnctn}
\end{eqnarray}
We introduce the following useful notation
\begin{eqnarray}
U^{(1,2)}_{ n \Omega q} (r,z) =  \phi^{(1,2)}_{n,\Omega,q} (r,z,\uparrow)\ , \\ 
L^{(1,2)}_{ n \Omega q} (r,z) =  \phi^{(1,2)}_{n,\Omega,q} (r,z,\downarrow) \ .
\end{eqnarray}
For axially symmetric systems, it is possible to eliminate the dependence on the
angle $\phi$, resulting in the \emph{reduced 2-D problem} in cylindrical coordinates
\cite{TOU02b}:

\begin{equation}
\left( 
\begin{array}{cc}
( h'^q -\lambda ) & \tilde h'^q \\
\tilde h'^q & - ( h'^q -\lambda ) 
\end{array}
\right)
\left(
\begin{array}{c}
\phi^{(1)}_{n,\Omega,q} \\  
\phi^{(2)}_{n,\Omega,q}
\end{array}
\right)
 = E_{n,\Omega,q}
\left( 
\begin{array}{c}
\phi^{(1)}_{n,\Omega,q} \\  
\phi^{(2)}_{n,\Omega,q}
\end{array}
\right)
\end{equation}
with 
\begin{eqnarray}
\left( 
\begin{array}{c}
\phi^{(1)}_{n,\Omega,q} \\ 
\phi^{(2)}_{n,\Omega,q} \\ 
\end{array} 
\right) 
&=&
\left( 
\begin{array}{c}
U^{(1)}_{n,\Omega,q} (r,z) \\ 
L^{(1)}_{n,\Omega,q} (r,z) \\
U^{(2)}_{n,\Omega,q} (r,z) \\ 
L^{(2)}_{n,\Omega,q} (r,z)
\end{array} 
\right) \ .
\end{eqnarray}

Here, quantities $\tilde{h'}$, $h'$, $U$ and $L$ are all functions of $(r,z)$ only.
This is the main mathematical structure that we implement in computational calculations.
For a given angular momentum projection quantum number $\Omega$, we solve
the eigenvalue problem to obtain energy 
eigenvalues $E_{n,\Omega,q}$ and eigenvectors $\psi_{n,\Omega,q}$
for the corresponding HFB quasiparticle states. 

From the definitions of the normal density and pairing density we
find the corresponding expressions in axial symmetry:
\begin{eqnarray}
  \rho_q(r,z) &=& \frac{1}{2 \pi} 
  \left(2 \sum_{\Omega>0}^{\Omega_{max}} \right) 
  \times \sum_{E_n>0}^{E_{max}} 
  \left[|U^{(2)}_{n \Omega q}(r,z)|^2  + |L^{(2)}_{n \Omega q}(r,z)|^2 \right] \\
  \tilde{\rho}_q(r,z) &=& - \frac{1}{2 \pi} 
  \left(2 \sum_{\Omega>0}^{\Omega_{max}} \right)  
  \times \sum_{E_n>0}^{E_{max}} 
  \left[U^{(2)}_{n \Omega q}(r,z) U^{(1)*}_{n \Omega q}(r,z)
  + L^{(2)}_{n \Omega q}(r,z) L^{(1)*}_{n \Omega q}(r,z) \right] \ .
\end{eqnarray}

\section{\label{sec:numerical}Lattice representation of spinor wavefunctions and Hamiltonian}

We solve the HFB eigenvalue problem by direct diagonalization on a 
two-dimensional grid $(r_\alpha,z_\beta)$, where $\alpha = 1,...,N_r$
and $\beta = 1,...,N_z$. The four components of the spinor wavefunction 
are represented on the two-dimensional lattice by an expansion in 
basis-spline functions $B_i (x)$ evaluated at the lattice support points. 
Further details about the basis-spline technique are given 
in Ref. \cite{Uma91,WO95}.

For the lattice representation of the Hamiltonian,
we use a hybrid method \cite{K96,KO96,Obe99} in which
derivative operators are constructed using the Galerkin method; this 
amounts to a global error reduction. Local potentials are
represented by the basis-spline collocation method (local error reduction).
The lattice representation transforms the differential operator
equation into a matrix form
\begin{equation}
\sum_{\nu=1}^N {\cal{H}}_{\mu}^{\ \nu} \psi^{\Omega}_{\nu} = 
            E^{\Omega}_{\mu} \psi^{\Omega}_{\mu} \ \ \ (\mu=1,...,N) \ ,
\end{equation}
The calculations use as a starting point the result of a {\it HF+BCS} previous 
calculation, which makes HFB converge substantially faster.
Since the problem is self-consistent we use an iterative method for the
solution. 
At every iteration the full HFB Hamiltonian is diagonalized. 
Due to the axial symmetry in the intrinsic frame, the diagonalization is
performed separately for each value of the angular momentum projection
quantum number $\Omega$ and for the two isospin projections $q=\pm \frac{1}{2}$.
Typically 20-30 iterations are sufficient for convergence at the level of one part
in $10^5$ for the total binding energy.

Note that in this lattice approach, the number of quasiparticle states
is determined by the dimensionality of the discrete HFB Hamiltonian
which is $N = (4 N_r N_z)^2$.

\section{Numerical results}

In the following we discuss some numerical results obtained with our new HFB-2D code.
First we present calculations for a light stable nucleus,
$^{22}_{10}$Ne. This nucleus was chosen because it has a large prolate
g.s. quadrupole deformation.
The calculation has been performed with the Skyrme SLy4 interaction in the p-h
channel and a pure delta pairing interaction (strength $V_0=-173 MeV fm^3$).
For the pairing forces of zero range employed here, one needs to introduce an energy
cut-off. In all of our calculations, we use a cut-off energy $E_{max}=60$ MeV in the
\emph{equivalent s.p. energy spectrum}, the same value utilized by Dobaczewski
et al. \cite{DN96} in his spherical 1-D calculations. 

\begin{figure}[h!]
\begin{center}
\includegraphics[scale=0.3]{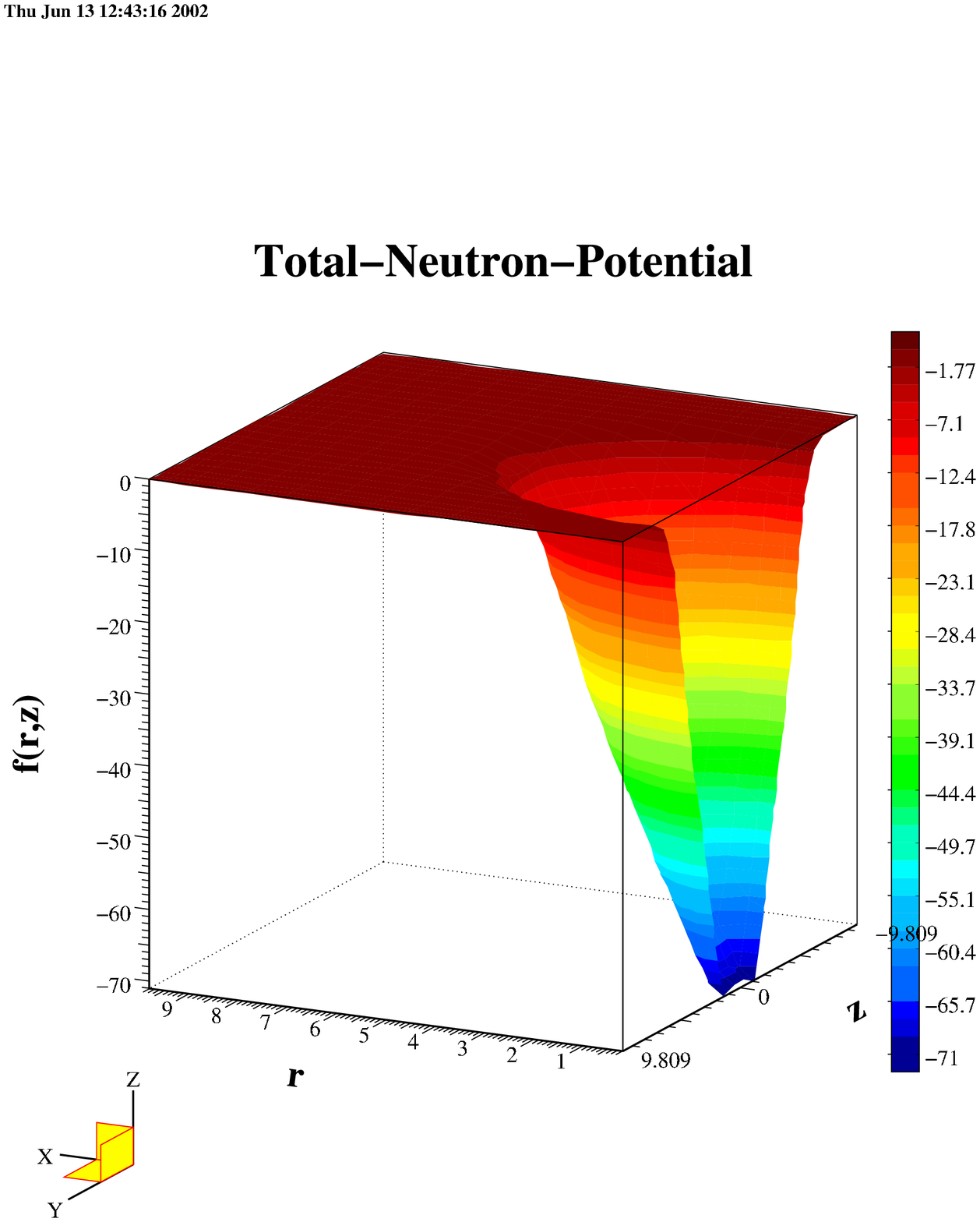}
\includegraphics[scale=0.3]{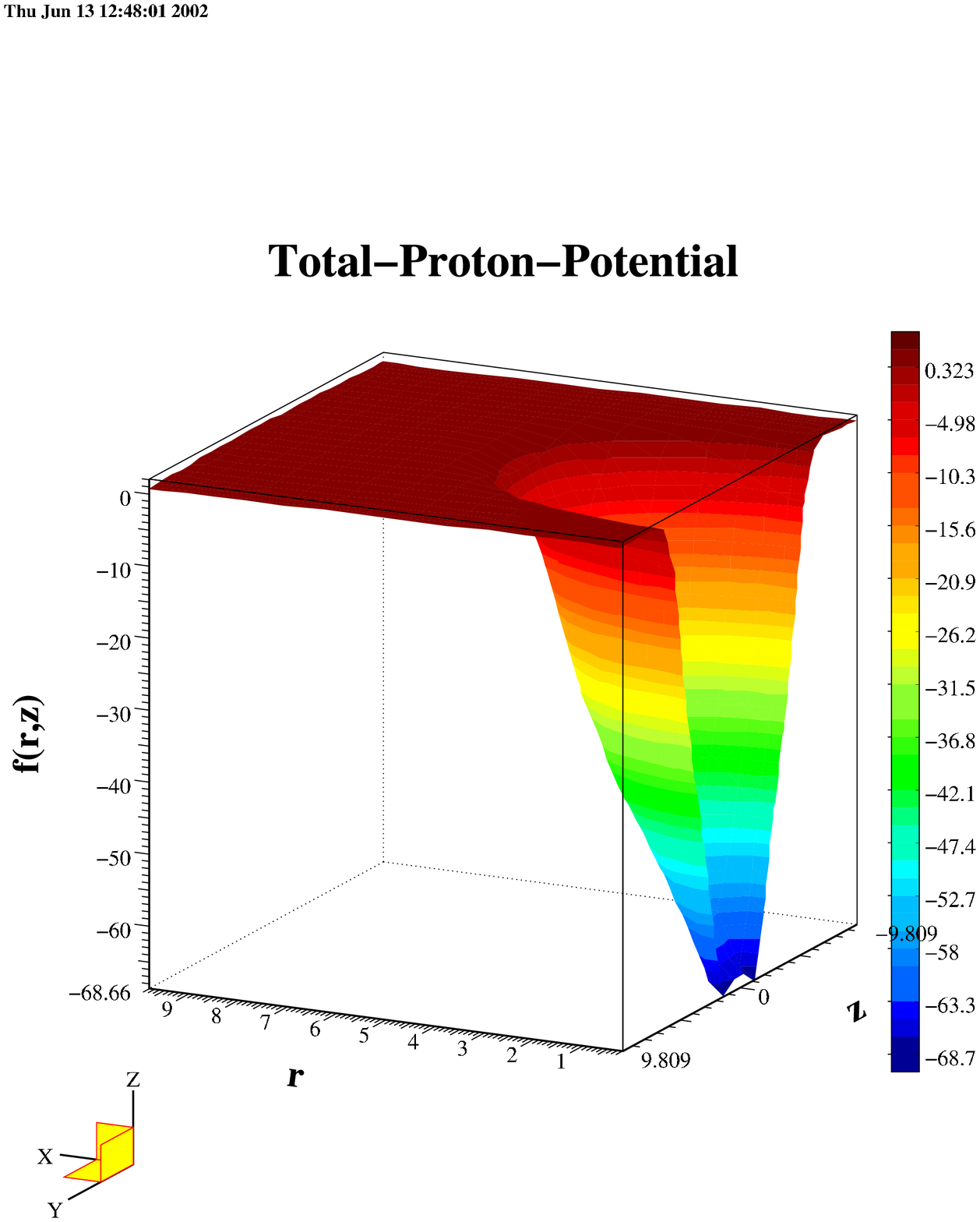}
\end{center}
\caption{$^{22}$Ne mean field potential for neutrons and protons}
\label{ne22_pot}
\end{figure}

\begin{figure}[h!]
\begin{center}
\includegraphics[scale=0.3]{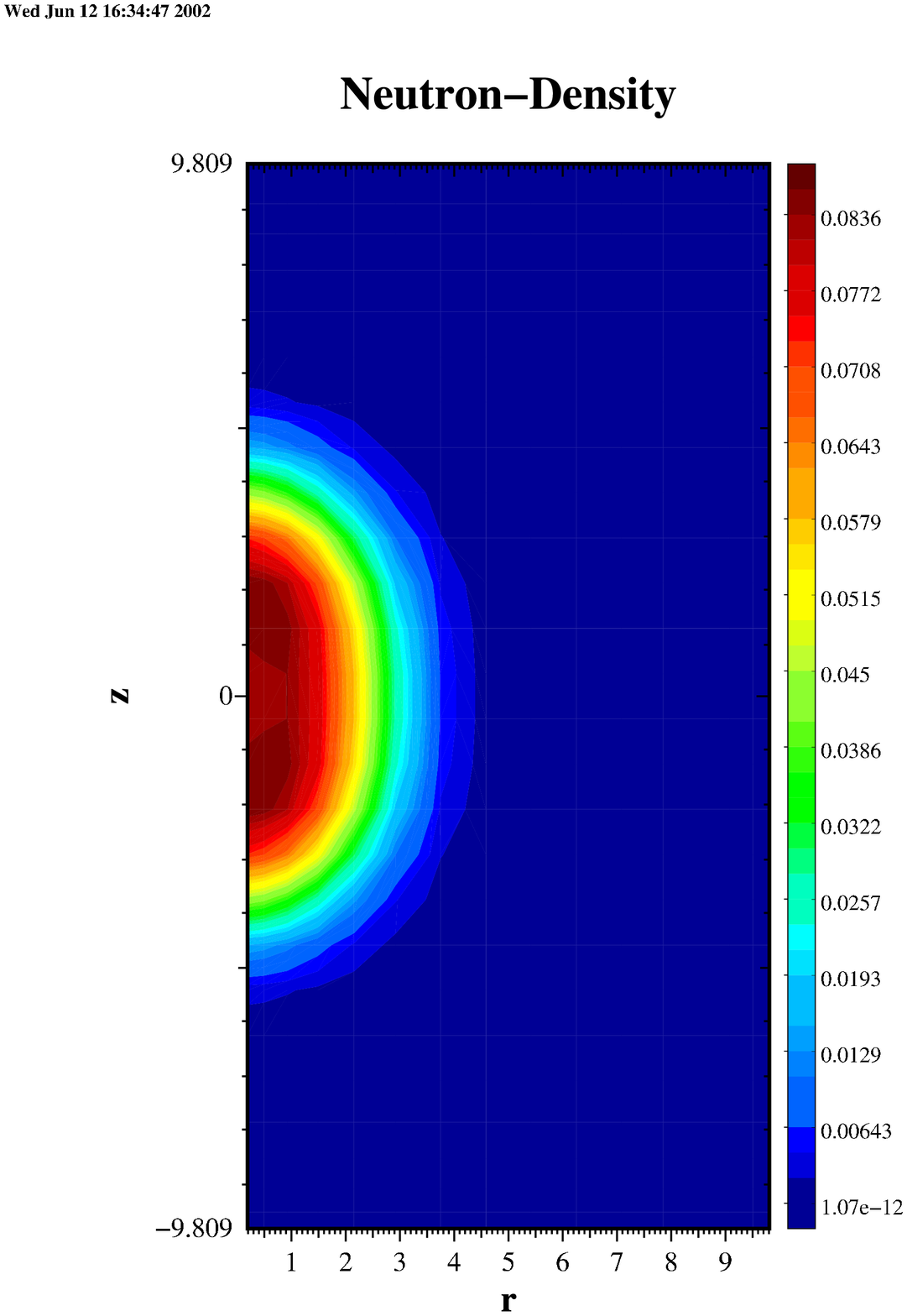}
\includegraphics[scale=0.3]{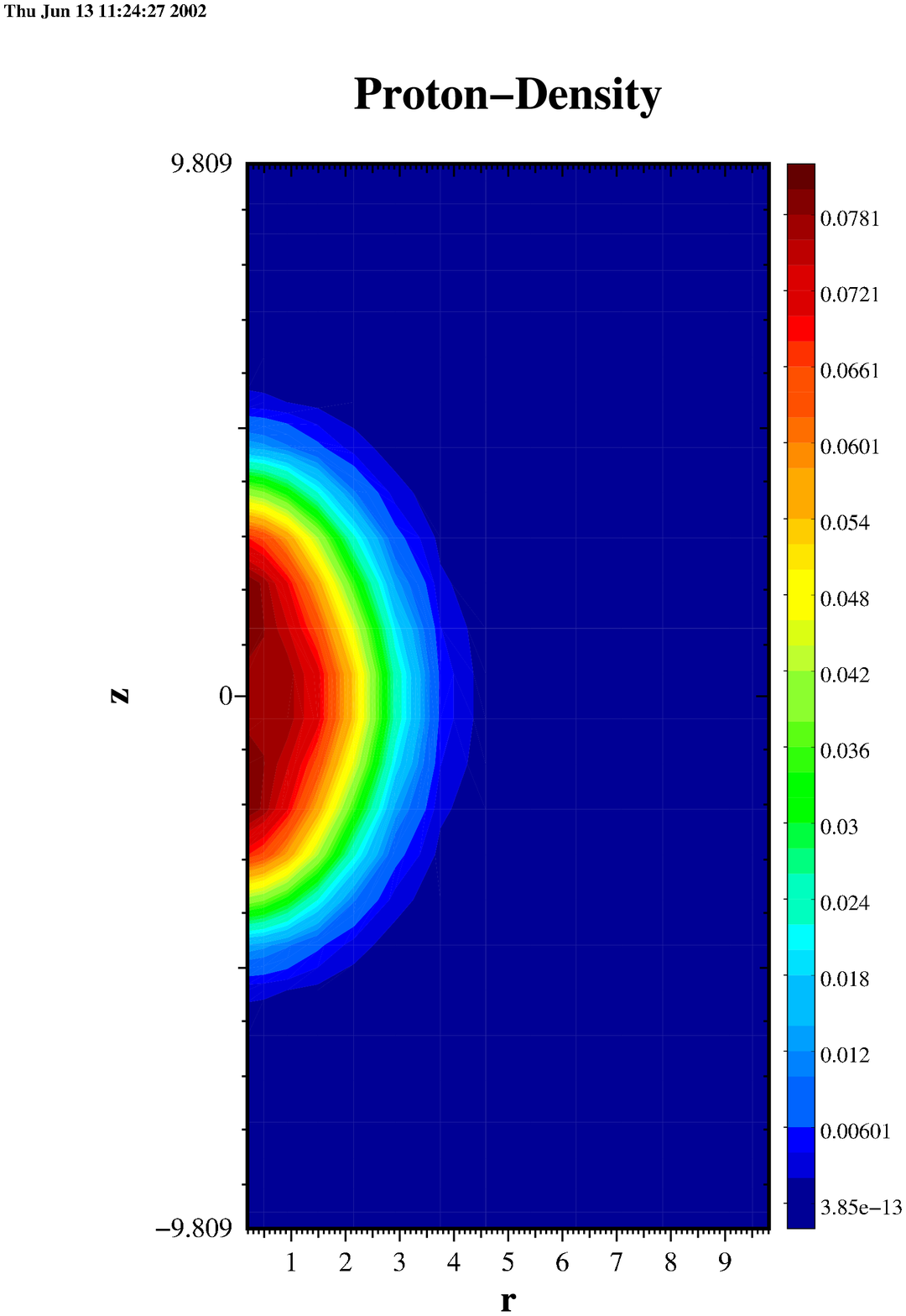}
\end{center}
\caption{Neutron and proton density distribution for the strongly deformed nucleus $^{22}$Ne}
\label{ne22_dens}
\end{figure}

\begin{figure}[h!]
\begin{center}
\includegraphics[scale=0.3]{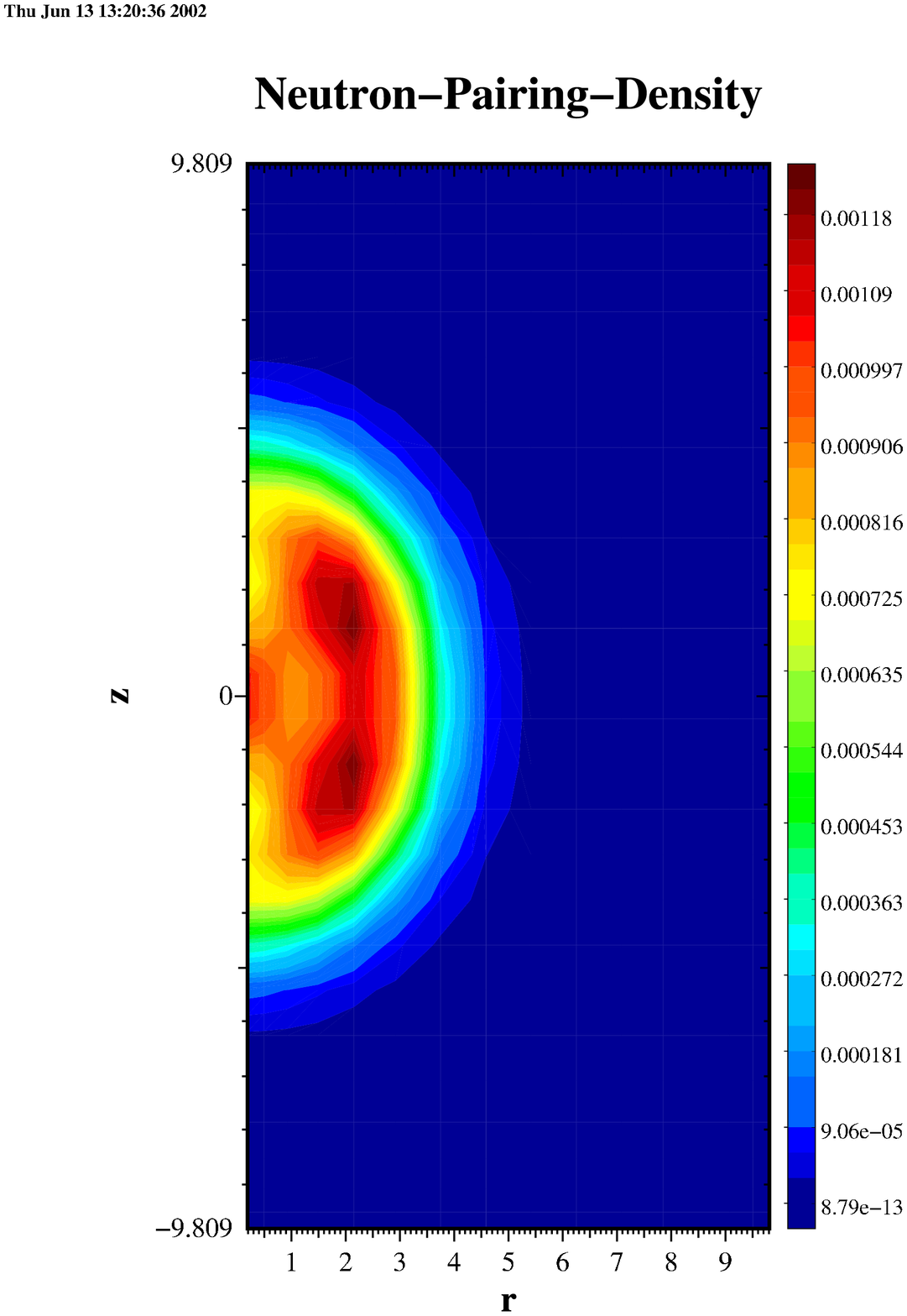}
\includegraphics[scale=0.3]{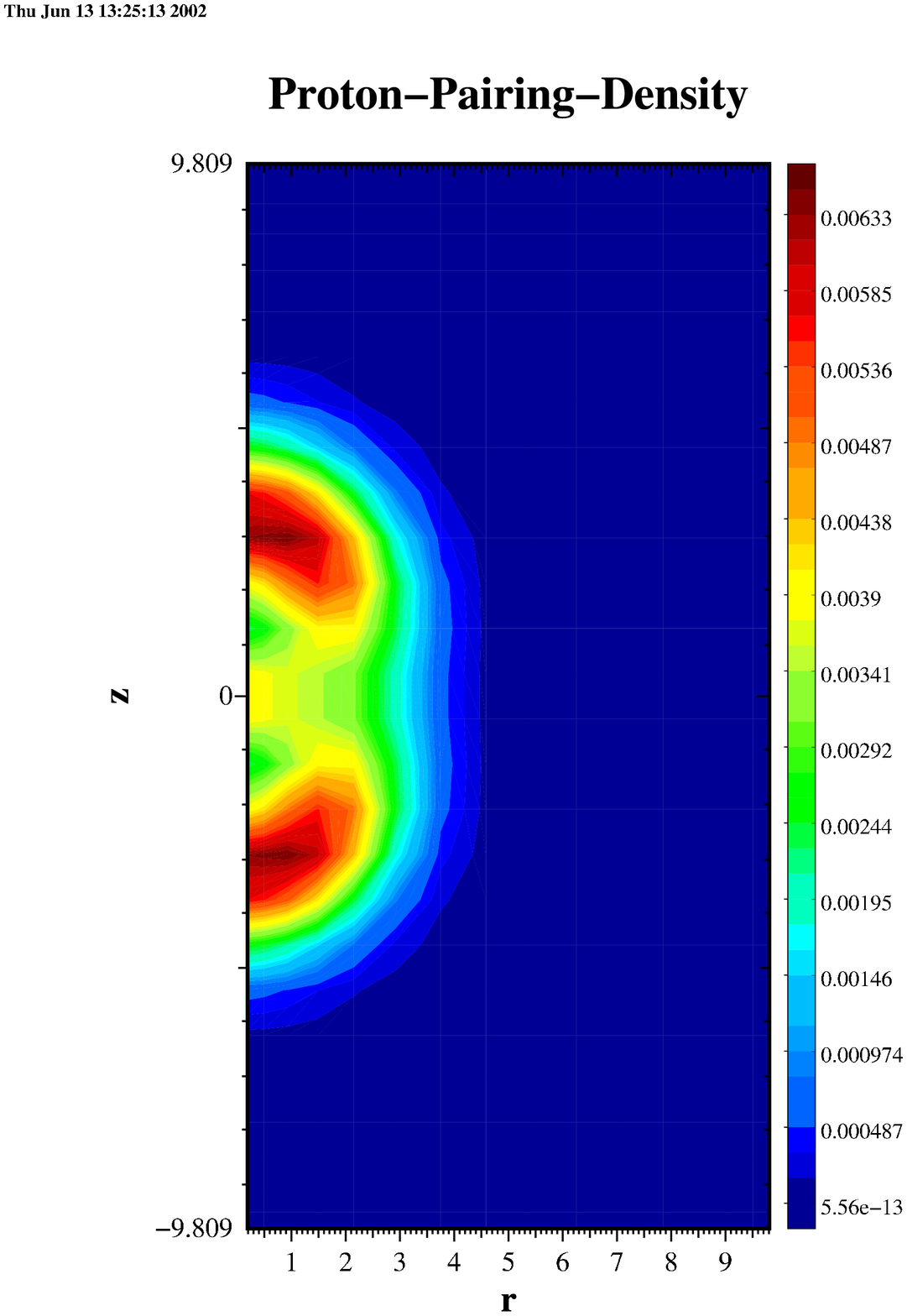}
\end{center}
\caption{Neutron and proton pairing density distribution for $^{22}$Ne}
\label{ne22_pairdens}
\end{figure}

Fig.~\ref{ne22_pot} shows the mean field potential $U_q({\bf r})$ for
neutrons and protons. Both mean fields are fairly similar, except that there is
an additional Coulomb contribution for the protons.
In Fig.~\ref{ne22_dens} we depict contour plots of the normal density for
neutrons and protons. The large prolate quadrupole deformation is clearly
visible in the density distributions.
Fig.~\ref{ne22_pairdens} shows the pairing density for
neutrons and protons. The square of the pairing density describes the probability
of finding a \emph{correlated} nucleon pair with opposite spin directions at position 
${\bf r}$ \cite{DN96}. Because this is a stable isotope, the pairing turns out to
be relatively weak. The shell structure of this light nucleus causes significant
differences in the neutron- vs. proton pairing.

We now present results for the n-rich isotope $^{22}O$ which is close
to the experimentally confirmed \cite{RIA1} dripline nucleus  $^{24}O$.
Because this nucleus turns out to be spherical in our 2-D calculations,
we can compare our results to the 1-D spherical code of Dobaczewski \cite{DF84}
and to the 2-D oscillator basis expansion method of Stoitsov et al. \cite{SD00}. 
Table \ref{table:comparison_o22} shows several observables for this nucleus:
the total binding energy, Fermi levels, pairing gaps and the r.m.s. radius.
Overall, the results of the axially symmetric code of the present work agree 
with the other two in all observables.

\begin{table}[h] 
\caption{\label{table:comparison_o22}
Comparison of several HFB calculations for $^{22} O$. In all cases,
the mean field is calculated with the SLY4 interaction, and the
pairing force is a pure delta interaction (strength $V_0=-218.5 MeV fm^3$)
corresponding to volume pairing. 
The axially symmetric calculations (2D) of this work used a box size 
$R = 10 fm$ with maximum $j_z = \frac{9}{2}$. 
The spherical calculation of Ref. \cite{Dpriv} was made with $R = 25 fm$ and a 
$j_{max}=\frac{21}{2}$.}
\begin{ruledtabular}
\begin{tabular}{ l c c c}
               &   1-D \cite{DN96,Dpriv} & 2-D (THO)\cite{SD00,Stpriv}  & 2-D (this work)  \\
\hline 
total binding energy (MeV)  & -164.60 & -164.52 & -164.64  \\
Fermi level, neutrons (MeV) &  -5.26  & -5.27   & -5.29  \\ 
Fermi level, protons (MeV)  &  -18.88 & -18.85  &  -18.08 \\ 
pairing gap, neutrons (MeV) & 1.42    & 1.41   &  1.36 \\ 
pairing gap, protons (MeV)  & 0.0     & 0.0    &  0.0 \\ 
r.m.s. radius  (fm)         &  2.92   & 2.92    &  2.92 \\ 
\end{tabular}
\end{ruledtabular}
\end{table}

\begin{figure}[h]
\begin{center}
\includegraphics[scale=0.3]{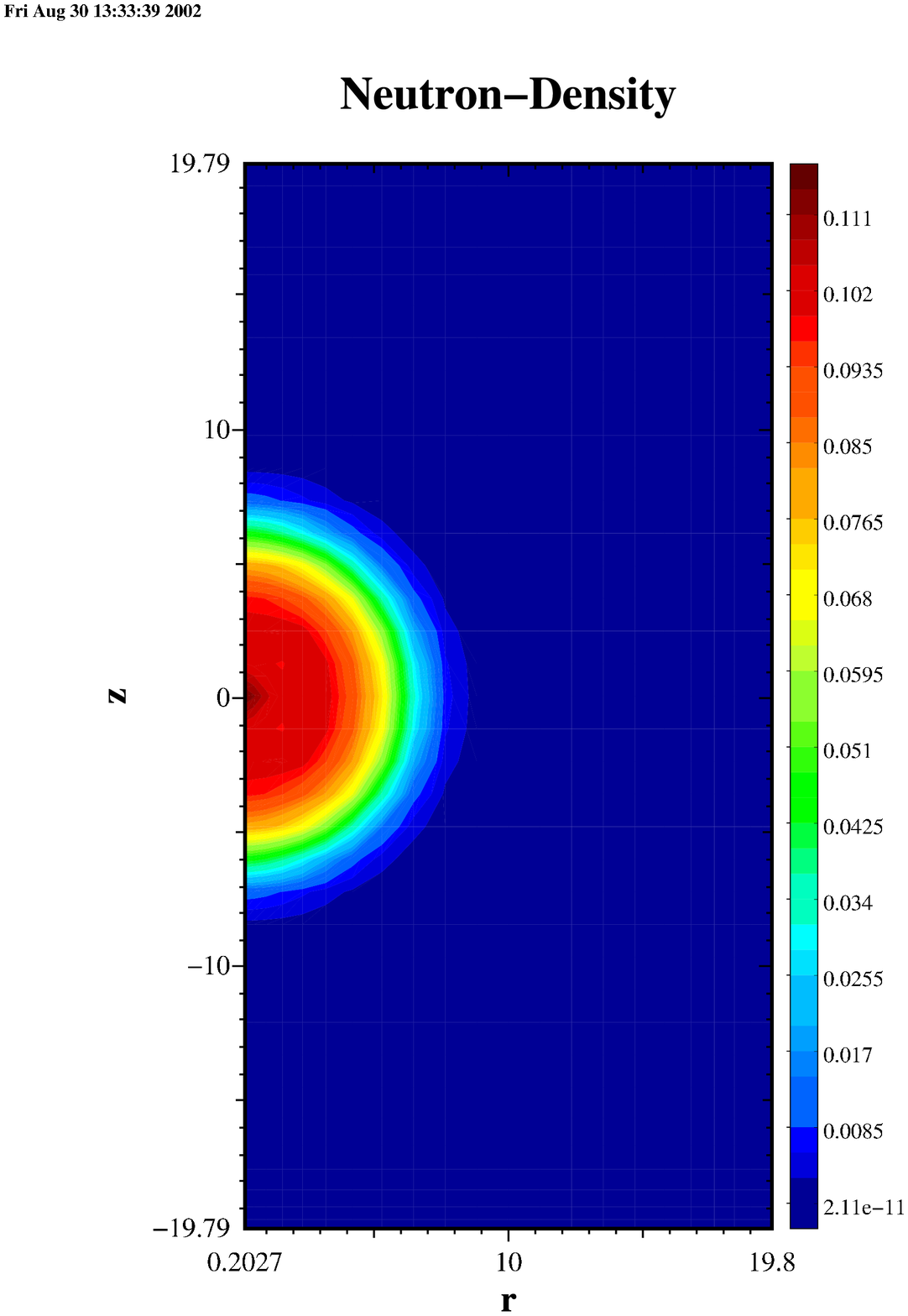}
\includegraphics[scale=0.3]{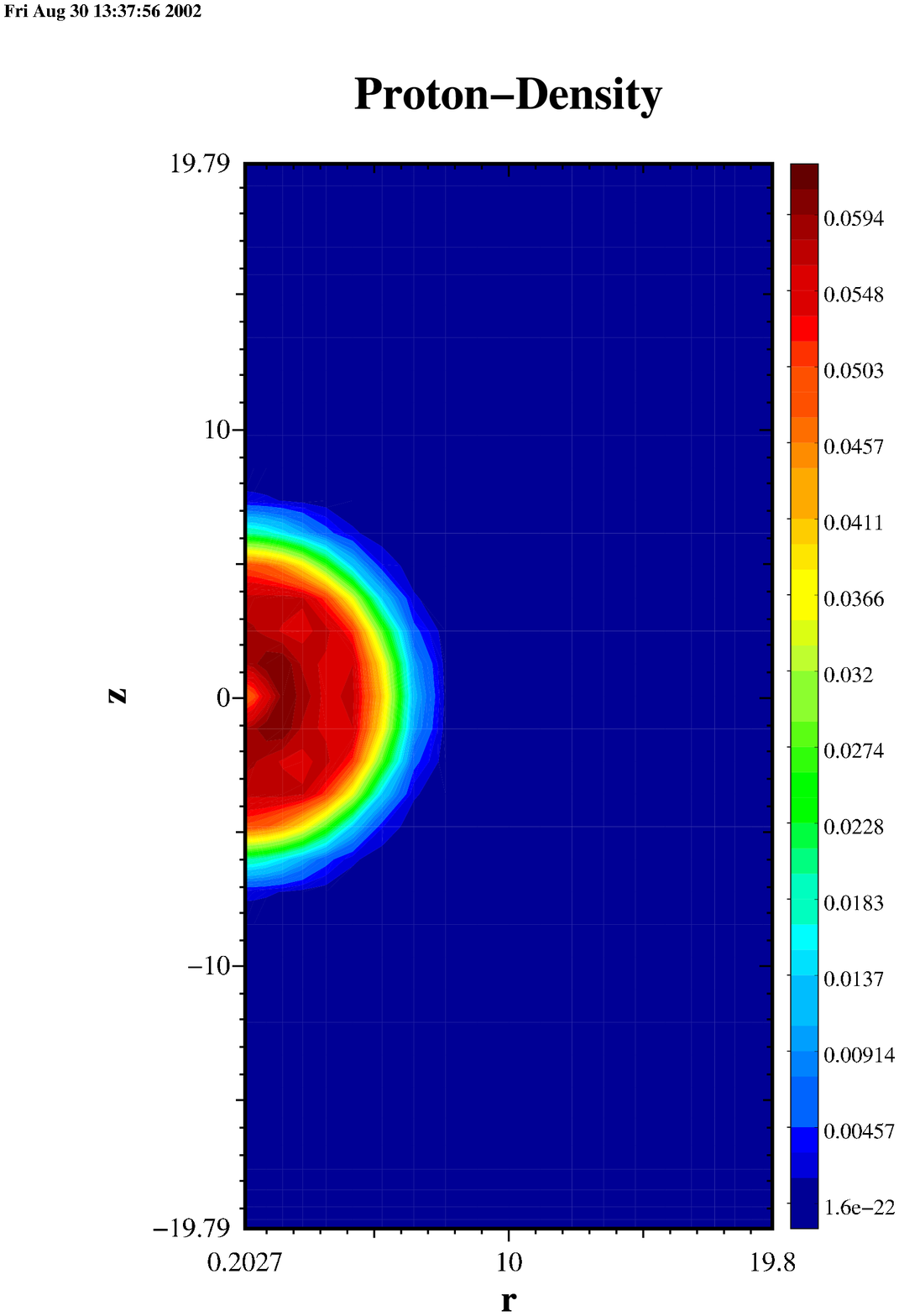}
\end{center}
\caption{Neutron and proton density distribution for $^{150}_{\ 50}$Sn, a nucleus near the
    two-neutron dripline}
\label{tin150_dens}
\end{figure}

\begin{figure}[h]
\begin{center}
\includegraphics[scale=0.3]{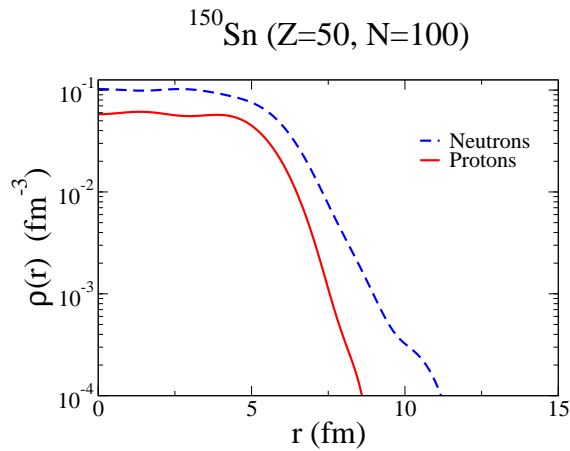}
\end{center}
\caption{Cut in radial direction of the neutron and proton density
in $^{150}_{\ 50}$Sn. The neutron skin is clearly visible.}
\label{tin150_denscut}
\end{figure}

We now turn our attention to the heavy neutron rich isotope $^{150}Sn$
$(Z=50,N=100)$ which has twice as many neutrons as protons. In
some phenomenological models, it is already close to the 2-neutron dripline.
Fig.~\ref{tin150_dens} shows contour plots of the normal density for
neutrons and protons. Because of the magic proton number and due to strong
pairing, the nuclear shape turns out to be spherical in our 2-D HFB code. In 
Fig.~\ref{tin150_denscut} we depict a cut of these
densities in radial direction which clearly exhibits the `neutron skin'
already seen in the 1-D calculations of Ref. \cite{DN96}. The
neutron skin represents a region of pure neutron matter which will allow
us to the study the properties of neutrons stars in future
laboratory experiments at RIA.

\begin{figure}[h]
\begin{center}
\includegraphics[scale=0.3]{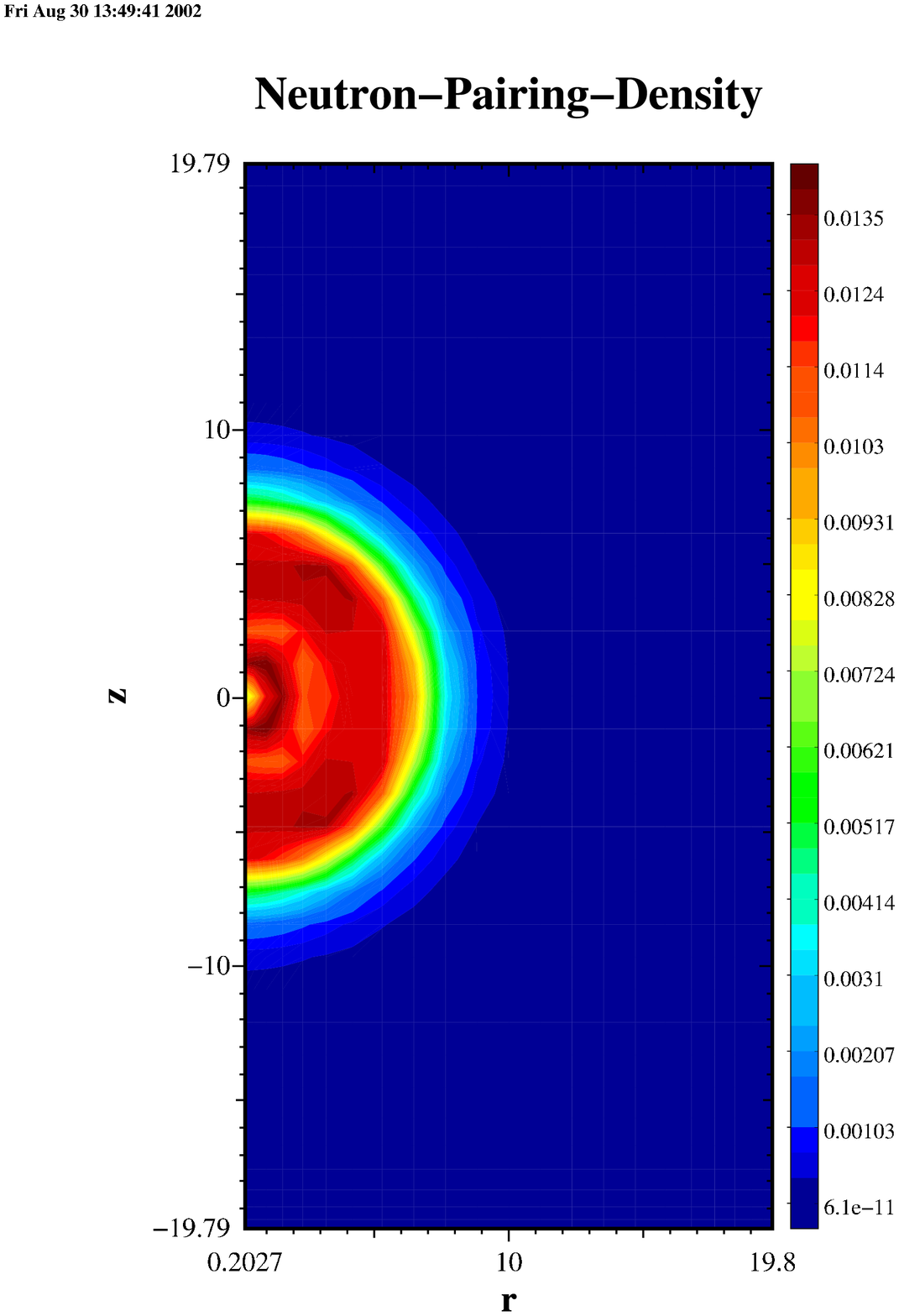}
\includegraphics[scale=0.3]{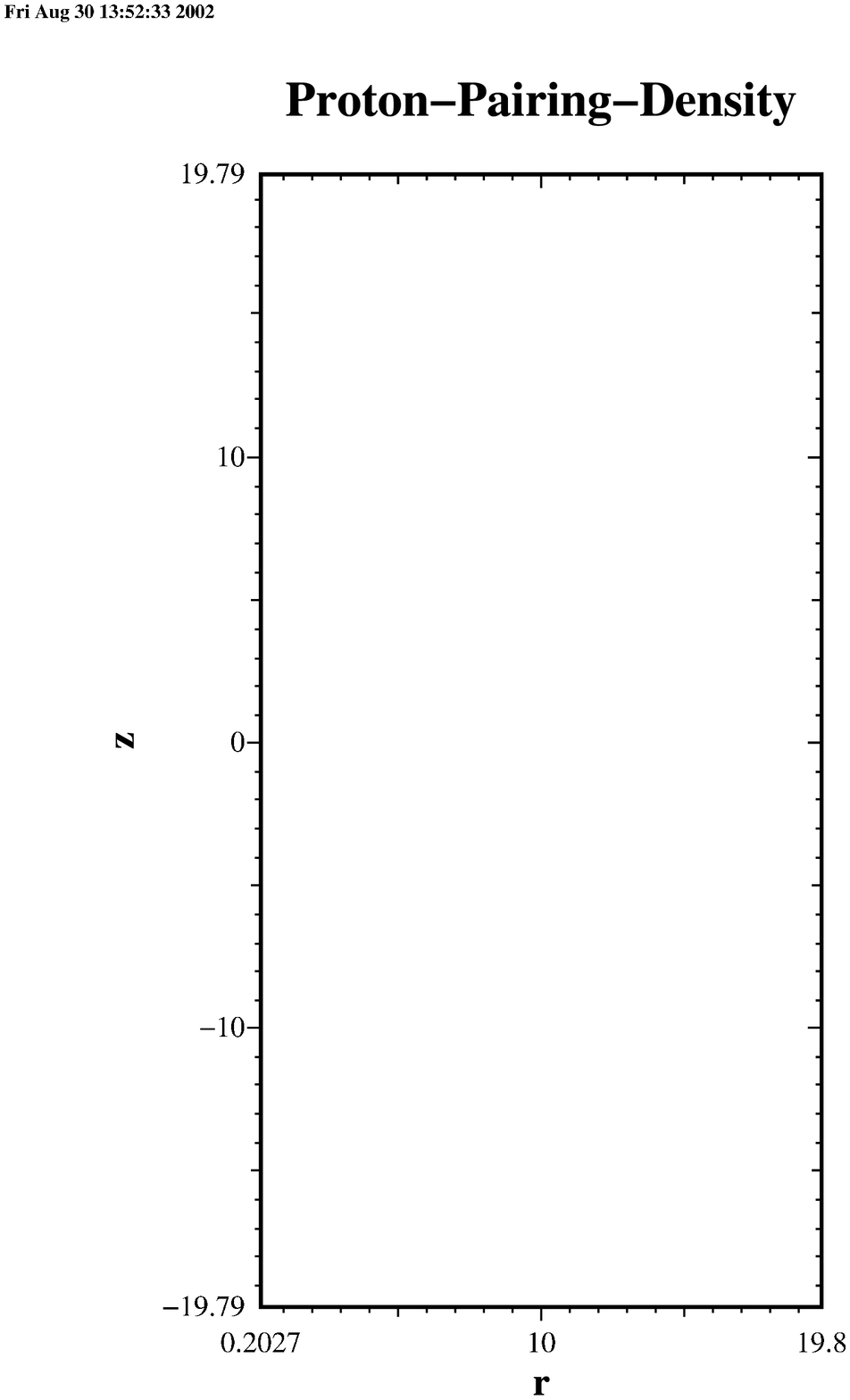}
\end{center}
\caption{Neutron and proton pairing density distribution for $^{150}_{\ 50}$Sn.
Because of the magic proton number $Z=50$ the proton pairing density is exactly
zero.}
\label{tin150_pairdens}
\end{figure}

In Table \ref{table:sn150} we compare again our 2-D calculations with
the 1-D radial results of Ref. \cite{DN96}.
In the 2-D calculations an approximately $3000\times 3000$ matrix was
diagonalized for each $j_z$ and isospin value, and for each major
HFB iteration. The full calculation required about 20 HFB iterations.
All observables agree quite well. The neutron Fermi level is less than
$1 MeV$ in this case which demonstrates clearly the proximity of
$^{150} Sn$ to the two-neutron drip line. 

\begin{table}[h]
\caption{\label{table:sn150}
Comparison of 1-D and 2-D calculations for $^{150} Sn$.
In both cases, the mean field is calculated with the SLY4 interaction, and the
pairing force is a pure delta interaction (strength $V_0=-170 MeV fm^3$)
corresponding to volume pairing. The 1-D calculations used a 
box size $R = 30$ with $j_{max}$ of $\frac{21}{2}$.
Calculations with our axially symmetric HFB 2-D code were made using a box size 
$R = 20 fm$ with maximum $j_z = \frac{13}{2}$.}
\begin{ruledtabular}
\begin{tabular}{ l c c}
               &   1-D \cite{DN96,Dpriv}  & 2-D (this work)  \\
\hline 
total binding energy (MeV)   &  -1129  &  -1130  \\
Fermi level, neutrons (MeV)  &  -0.96  &  -0.94  \\ 
Fermi level, protons  (MeV)  &  -17.54 &  -17.34 \\ 
pairing gap, neutrons (MeV)  &   1.02  &   0.97  \\ 
pairing gap, protons (MeV)   &   0.00  &   0.00  \\
r.m.s. radius  (fm)          &   5.126 &  5.132 \\ 
\end{tabular}
\end{ruledtabular}
\end{table}

\section{\label{sec:conclusions}Summary and Conclusions}

Our goal for the near future is to investigate several isotope chains, in particular
deformed nuclei, and to calculate observables which are important for the
physics near the drip lines, i.e. binding energies, neutron and proton
separation energies, pairing gaps, particle densities and pairing
densities, rms radii, and electric or magnetic moments. We plan to utilize
a variety of Skyrme parameterizations for the mean field, and both
volume and surface pairing forces.
As more data from existing RIB facilities become available, it is likely that 
it will become necessary to develop new effective N-N
interactions to describe these exotic nuclei. Furthermore, our 2-D HFB code
results may be used as input into coordinate-space based QRPA calculations
to investigate collective excited states (surface vibrations and giant
resonances). 

We will also study alternative numerical techniques to speed
up our 2-D HFB code, in particular damping methods which we have utilized
successfully in solving the Dirac equation \cite{WO95}. Provided that the
damping method can be successfully implemented in 2-D, we will attempt
to solve the 3-D HFB problem in Cartesian coordinates. This will certainly be very difficult,
but it is worth trying: in the 1996 DOE/NSAC Long Range Plan,
unrestricted HFB theory on the lattice has been described as a computational
Grand Challenge project in nuclear physics.

\begin{acknowledgments}
This work is supported by the U.S. Department of Energy under grant No.
DE-FG02-96ER40963 with Vanderbilt University. Some of the numerical calculations
were carried out on supercomputers at the National Energy Research
Scientific Computing Center (NERSC). We also like to acknowledge many
fruitful discussions with W. Nazarewicz and M. Stoitsov (ORNL) and with
J. Dobaczewski (Warsaw).
\end{acknowledgments}


\end{document}